\documentclass[runningheads]{llncs}
\usepackage[T1]{fontenc}
\usepackage{amsmath}
\usepackage{amssymb}

\usepackage{fancyvrb}       

\usepackage[most]{tcolorbox}
\usepackage{inconsolata}
\usepackage{xcolor}
\usepackage{tikz}
\usepackage{graphicx}
\usepackage{everysel}
\usepackage{float}
\usepackage{caption}
\usepackage{gensymb}

\EverySelectfont{%
\fontdimen2\font=0.4em
\fontdimen3\font=0.2em
\fontdimen4\font=0.1em
\fontdimen7\font=0.1em
}

\begin{document}

\title{Towards a constructive formalization of Perfect Graph Theorems} 
\author{Abhishek Kr Singh\inst{1} \and Raja Natarajan\inst{2}}

\institute{Tata Institute of Fundamental Research, Mumbai\\ \email{abhishek.singh@tifr.res.in} \and 
Tata Institute of Fundamental Research, Mumbai\\  \email{raja@tifr.res.in}}

\newcommand{\tw}{\texttt}

\definecolor{vlgray}{gray}{1}
\def \d {1}
\tikzset{
vertR/.style={scale=0.5,circle, draw=red!150, fill=red!100,thick},
vertB/.style={scale=0.5,circle, draw=blue!150, fill=blue!100,thick},
vertG/.style={scale=0.5,circle, draw=green!150, fill=green!100,thick},
vertY/.style={scale=0.5,circle, draw=yellow!250, fill=yellow!200,thick},
vertBlack/.style={scale=0.5,circle, draw=black!150, fill=black!100,thick}, 
vertC/.style={scale=0.5,circle,draw=cyan!150,fill=cyan!100,thick}}

\newcommand{\UTriangle}
{
\path(0,0) coordinate (Origin);
\path (Origin)+ (0:0) coordinate(A) +(0: \d) coordinate(B) + (60: \d) coordinate(C);
\draw (A)--(B)--(C)--(A);
\node (a) at (A) [vertR] {};
\node (b) at (B) [vertB] {};
\node (c) at (C) [vertG] {};
}

%
\newcommand{\USquare}
{
\path(2,0) coordinate (Origin);
\path(Origin) ++(0:0) coordinate(A) ++(0:\d) coordinate(B) ++ (90:\d) coordinate(C) ++ (180:\d) coordinate(D);
\draw (A)--(B)--(C)--(D)--(A);
\node (a) at (A)[vertR]{};
\node (b) at (B)[vertB]{};
\node (c) at (C)[vertR]{};
\node (d) at (D)[vertB]{};
}

%
%
\newcommand{\URectangle}
{
\path(0,0) coordinate (Origin);
\path(Origin) ++(0:0) coordinate(A) ++(0:\d+1) coordinate(B) ++ (90:\d) coordinate(C) ++ (180:\d+1) coordinate(D);
\draw (A)--(B)--(C)--(D)--(A);
\node (a) at (A)[vertR]{};
\node (b) at (B)[vertB]{};
\node (c) at (C)[vertR]{};
\node (d) at (D)[vertB]{};
}

\newcommand{\UPentagon}
{
\path(0,0) coordinate (Origin);
\path (Origin) ++ (0:0) coordinate(Current);
\foreach \i in{0,1,2,...,4}
{
\path (Current)++(0:0) coordinate(V\i);
\path (Current)++(\i*72: \d) coordinate(Current);
}
\draw (V0)--(V1)--(V2)--(V3)--(V4)--(V0);
\node (v0) at (V0) [vertR]{};
\node (v1) at (V1) [vertB]{};
\node (v2) at (V2) [vertR]{};
\node (v3) at (V3) [vertB]{};
\node (v4) at (V4) [vertG]{};
}

\newcommand{\UStar}
{
\path(2,0) coordinate (Origin);
\path (Origin) ++ (0:0) coordinate(A) ++ (0:\d) coordinate(B) ++ (72:\d) coordinate(C) ++ (2*72:\d) coordinate(D) ++ (3*72:\d) coordinate(E);

\draw (A)--(C);
\draw (B)--(D);
\draw (C)--(E);
\draw (B)--(E);
\draw (A)--(D);
\node (a)[vertR] at (A){};
\node (b)[vertB] at (B){};
\node (c)[vertB] at (C){};
\node (d)[vertG] at (D){};
\node (e)[vertR] at (E){};
}

\newcommand{\UDoublepentagon}
{
\path(5,0) coordinate (Origin);
\path (Origin) ++ (0:0) coordinate(Current);
\foreach \i in{0,1,2,...,4}
{
\path (Current)++(0:0) coordinate(V\i);
\path (Current)++(\i*72: \d) coordinate(Current);
}
\draw (V0)--(V1)--(V2)--(V3)--(V4)--(V0);
\path(8,0)coordinate(Origin);
\path (Origin) ++ (0:0) coordinate(Current);
\foreach \i in{0,1,2,...,4}
{
\path (Current)++(0:0) coordinate(U\i);
\path (Current)++(\i*72: \d) coordinate(Current);
}
\draw (U0)--(U1)--(U2)--(U3)--(U4)--(U0);
\draw (V2)--(U4)--(V3)--(U3)--(V2);
\node (v0) at (V0) [vertR]{};
\node (v1) at (V1) [vertB]{};
\node (v2) at (V2) [vertR]{};
\node (v3) at (V3) [vertB]{};
\node (v4) at (V4) [vertG]{};
\node (u0) at (U0) [vertC]{};
\node (u1) at (U1) [vertY]{};
\node (u2) at (U2) [vertC]{};
\node (u3) at (U3) [vertY]{};
\node (u4) at (U4) [vertBlack]{};
}


\newcommand{\UBipartitegraph}
{
\path(4,0)coordinate(Origin);
\def \dy  {\d/2}
\path(Origin) ++ (0,0) coordinate(l1) ++ (0,\dy) coordinate(l2) ++ (0,\dy) coordinate(l3) ++ (0,\dy) coordinate(l4) ++ (1,0) coordinate(r4) ++ (0,-\dy) coordinate(r3) ++ (0,-\dy) coordinate(r2) ++ (0,-\dy) coordinate(r1);
\draw  (r1)--(l2);
\draw (l3)--(r3);
\draw (l4)--(r2);
\draw (r4)--(l3);
\node ()[vertR] at (l1){};
\node ()[vertR] at (l2){};
\node ()[vertR] at (l3){};
\node ()[vertR] at (l4){};
\node ()[vertB] at (r1){};
\node ()[vertB] at (r2){};
\node ()[vertB] at (r3){};
\node ()[vertB] at (r4){};
}


\newcommand{\ULine}
{
\path (0,0) coordinate(Origin);
\path(Origin) +(0,0) coordinate(v1)+ (\d,0)coordinate(v2);
\draw(v1)--(v2);
\node ()[vertR] at(v1){};
\node ()[vertB] at(v2){};
}

\newcommand{\USkewedpentagon}
{
\path(0,0) coordinate(Origin);
\path(Origin) ++ (0:0) coordinate(A) ++ (0:\d) coordinate(B) ++ (1*72:\d) coordinate(C) ++ (2*72:\d) coordinate(D) ++ (3*72:\d) coordinate(E);
\draw (A)--(C)--(D)--(E)--(B)--(A);
\node() [vertR] at (A){};
\node() [vertG] at (B){};
\node() [vertB] at (C){};
\node() [vertR] at (D){};
\node() [vertB] at (E){};
}

\newcommand{\UExtendedpentagon}
{
\path(0,0) coordinate (Origin);
\path (Origin) ++ (0:0) coordinate(Current);
\foreach \i in{0,1,2,...,4}
{
\path (Current)++(0:0) coordinate(V\i);
\path (Current)++(\i*72: \d) coordinate(Current);
}
\path (V2) ++ (25:2*\d/3) coordinate (V5);
\draw (V0)--(V1)--(V2)--(V3)--(V4)--(V0);
\draw (V1)--(V5)--(V3);
\draw (V2)-- (V5);
\node (v0) at (V0) [vertR]{};
\node (v1) at (V1) [vertB]{};
\node (v2) at (V2) [vertR]{};
\node (v3) at (V3) [vertB]{};
\node (v4) at (V4) [vertG]{};
\node (v5) at (V5) [vertG]{};
}
\newcommand{\LGraph}
{ 
\path [fill=gray!10] (-1.5,-1) rectangle (1.7,3);
\path(0,0) coordinate(Origin);
\path(Origin) ++ (0:0) coordinate(V2) ++ (5:\d) coordinate(V3) ++ (110:2*\d) coordinate(A) ++ (225:1.25*\d) coordinate(V1);
\draw (V2)--(A)--(V1);
\draw (V3)--(A);
\node (v2) at (V2) [vertBlack,label={below:{$v_{2}$}}]{};
\node (v3) at (V3) [vertBlack,label={below:{$v_{3}$}}]{};
\node (a) at (A) [vertBlack,label={above:{$a$}}]{};
\node (v1) at (V1) [vertBlack,label={below:{$v_{1}$}}]{};
\path (A) ++ (160:1.5) coordinate (G);
\node (g) at (G) []{$G$};

\path(5*\d,0)coordinate(Origin);
\path [fill=gray!10] (3.7,-1) rectangle (7,3);
\path(Origin) ++ (0:0) coordinate(V2) ++ (5:\d) coordinate(V3) ++ (110:2*\d) coordinate(A) ++ (225:1.25*\d) coordinate(V1);
\path (V3) ++ (86:1.5*\d) coordinate (A'); 
\draw (V2)--(A)--(V1);
\draw (V3)--(A);
\draw [thin,dashed] (V2)--(A')--(V1);
\draw [thin,dashed] (V3)--(A')--(A);
\node (v2) at (V2) [vertBlack,label={below:{$v_{2}$}}]{};
\node (v3) at (V3) [vertBlack,label={below:{$v_{3}$}}]{};
\node (a') at (A') [scale=0.5,circle, draw=black!50, fill=black!20,thick,label={above:{$a^{\prime}$}}]{};
\node (a) at (A) [vertBlack,label={above:{$a$}}]{};
\node (v1) at (V1) [vertBlack,label={below:{$v_{1}$}}]{};

\path (A) ++ (160:1.3) coordinate (G');
\node (g') at (G') []{$G'$};
}
\newcommand{\LPolygonABCD}
{
\path(0,0)coordinate(origin);
\path(origin) ++ (0:0) coordinate(D) ++ (0:1.5*\d) coordinate(C) ++ (90:\d) coordinate(B);
\path(origin) ++ (90:1.5*\d) coordinate(A);
\draw (D)--(C)--(B)--(A)--(D);
\node (d) at (D) [vertBlack,label={below left:{$d$}}]{};
\node (c) at (C) [vertBlack,label={below right:{$c$}}]{};
\node (b) at (B) [vertBlack,label={above right:{$b$}}]{};
\node (a) at (A) [vertBlack,label={above left:{$a$}}]{};

\path(origin) ++ (320:\d) coordinate(G1);
\node (g1) at (G1) [label={below : {$G_{1}$}}]{};
}

\newcommand{\LPolygonVaBCD}
{
\path (2.5,0) coordinate (Origin);
\path(Origin) ++ (0:0) coordinate(D) ++ (0:1.5*\d) coordinate(C) ++ (90:\d) coordinate(B);
\path(Origin) ++ (90:1.5*\d) coordinate(A);
\path(A) ++ (30:0.25*\d) coordinate (A2);
\path(A) ++ (210:0.25*\d) coordinate (A1);
\foreach \i in{1,2}
{
\draw [gray!40](A\i)--(D); 
}
\foreach \i in{1,2}
{
\draw [gray!40](A\i)--(B);  
}
\node (a1) at (A1) [vertBlack,label={above : {$V_{a}$}}]{};
\node (a2) at (A2) [vertBlack]{};
\node (b) at (B) [vertBlack, label={above : {$b$}}]{};
\node (c) at (C) [vertBlack, label={below : {$c$}}]{};
\node (d) at (D) [vertBlack, label={below : {$d$}}]{};
\draw (A1)--(A2);
\draw (D)--(C)--(B);
\path(Origin) ++ (320:\d) coordinate(G2);
\node (g2) at (G2) [label={below : {$G_{2}$}}]{};
}

%
%
\newcommand{\LPolygonVaVbCD}
{
\path(5,0)coordinate(origin);
\path(origin) ++ (0:0) coordinate(D) ++ (0:1.5*\d) coordinate(C) ++ (90:\d) coordinate(B);
\path(origin) ++ (90:1.5*\d) coordinate(A);
\path(A) ++ (30:0.25*\d) coordinate (A2);
\path(A) ++ (210:0.25*\d) coordinate (A1);
\path(B) ++ (0:0) coordinate (B1) ++ (20:0.5*\d) coordinate (B2) ++ (140:0.5*\d) coordinate (B3);
\foreach \i in{1,2}
{
\foreach \j in {1,2,3}
{ \draw [gray!40](A\i)--(B\j); } 
}
\foreach \i in{1,2,3}
{
\draw [gray!40](B\i)--(C); 
}
\foreach \i in{1,2}
{
\draw [gray!40](A\i)--(D); 
}
\node (a1) at (A1) [vertBlack,label={above : {$V_{a}$}}]{};
\node (a2) at (A2) [vertBlack]{};
\node (b1) at (B1) [vertBlack]{};
\node (b2) at (B2) [vertBlack, label={above : {$V_{b}$}}]{};
\node (b3) at (B3) [vertBlack]{};
\node (c) at (C) [vertBlack, label={below : {$c$}}]{};
\node (d) at (D) [vertBlack, label={below : {$d$}}]{};
\draw (A1)--(A2);
\draw (B1)--(B2)--(B3)--(B1);
\draw (C)--(D);
\path(origin) ++ (320:\d) coordinate(G3);
\node (g3) at (G3) [label={below : {$G_{3}$}}]{};
}

\newcommand{\LExtendedpolygon}
{
\path(8,0)coordinate(origin);
\path(origin) ++ (0:0) coordinate(D) ++ (0:1.5*\d) coordinate(C) ++ (90:\d) coordinate(B);
\path(origin) ++ (90:1.5*\d) coordinate(A);
\path(A) ++ (30:0.25*\d) coordinate (A2);
\path(A) ++ (210:0.25*\d) coordinate (A1);
\path(D) ++ (270:0.2) coordinate (D1);
\path(C) ++ (0,0) coordinate (C1) ++ (0:0.5*\d) coordinate (C2) ++ (270:0.5*\d) coordinate (C3) ++ (180:0.5*\d) coordinate (C4);
\path(B) ++ (0:0) coordinate (B1) ++ (20:0.5*\d) coordinate (B2) ++ (140:0.5*\d) coordinate (B3);

\foreach \i in{1,2}
{
\foreach \j in {1,2,3}
{ \draw [gray!40](A\i)--(B\j); } 
}
\foreach \i in{1,2,3}
{
\foreach \j in {1,2,3,4}
{ \draw [gray!40](B\i)--(C\j); } 
}
\foreach \i in{1}
{
\foreach \j in {1,2,3,4}
{ \draw [gray!40](D\i)--(C\j); } 
}
\foreach \i in{1,2}
{
\foreach \j in {1}
{ \draw [gray!40](A\i)--(D\j); } 
}

\node (a1) at (A1) [vertBlack,label={above : {$V_{a}$}}]{};
\node (a2) at (A2) [vertBlack]{};
\node (b1) at (B1) [vertBlack]{};
\node (b2) at (B2) [vertBlack, label={above : {$V_{b}$}}]{};
\node (b3) at (B3) [vertBlack]{};
\node (c1) at (C1) [vertBlack]{};
\node (c2) at (C2) [vertBlack]{};
\node (c3) at (C3) [vertBlack, label={right : {$V_{c}$}}]{};
\node (c4) at (C4) [vertBlack]{};
\node (d1) at (D1) [vertBlack, label={below : {$V_{d}$}}]{};
\draw (A1)--(A2);
\draw (B1)--(B2)--(B3)--(B1);
\draw (C1)--(C2)--(C3)--(C4)--(C1);
\path(origin) ++ (320:\d) coordinate(G4);
\node (g4) at (G4) [label={below : {$G_{4}$}}]{};
}


\newcommand{\LPentagon}
{
\path(0,0) coordinate (Origin);
\path (Origin) ++ (0:\d) coordinate(Current);
\foreach \i in{1,2,...,5}
{
\path (Current)++(0:0) coordinate(V\i);
\path (Current)++(\i*72: \d) coordinate(Current);
}
\draw (V1)--(V2)--(V3)--(V4)--(V5)--(V1);
\node (v1) at (V1) [vertB,label=below:{$v_{1}$}]{};
\node (v2) at (V2) [vertR,label=right:{$v_{2}$}]{};
\node (v3) at (V3) [vertB,label=above:{$v_{3}$}]{};
\node (v4) at (V4) [vertG,label=left:{$v_{4}$}]{};
\node (v5) at (V5) [vertR,label=below:{$v_{5}$}]{};
\path(Origin) ++ (300:\d) coordinate(G1);
\node (g1) at (G1) []{$G_{1}$};
}
%
%

\newcommand{\LExtendedpentagon}
{
\path(3,0) coordinate (Origin);
\path (Origin) ++ (0:\d) coordinate(Current);
\foreach \i in{1,2,...,5}
{
\path (Current)++(0:0) coordinate(V\i);
\path (Current)++(\i*72: \d) coordinate(Current);
}
\path(V4) ++ (340:0.3*\d) coordinate(V4');
\draw (V1)--(V2)--(V3)--(V4)--(V5)--(V1);
\draw (V3)--(V4')--(V5);
\draw (V4)--(V4');
\node (v1) at (V1) [vertB,label=below:{$v_{1}$}]{};
\node (v2) at (V2) [vertG,label=right:{$v_{2}$}]{};
\node (v3) at (V3) [vertR,label=above:{$v_{3}$}]{};
\node (v4) at (V4) [vertB,label=left:{$v_{4}$}]{};
\node (v5) at (V5) [vertR,label=below:{$v_{5}$}]{};
\node (v4') at (V4') [vertG,label=right:{$v_{4}^{\prime}$}]{};
\path(Origin) ++ (300:\d) coordinate(G2);
\node (g2) at (G2) []{$G_{2}$};
}
%
%

\newcommand{\LDextendedpentagon}
{
\path(6,0) coordinate (Origin);
\path (Origin) ++ (0:\d) coordinate(Current);
\foreach \i in{1,2,...,5}
{
\path (Current)++(0:0) coordinate(V\i);
\path (Current)++(\i*72: \d) coordinate(Current);
}
\path(V4) ++ (340:0.3*\d) coordinate(V4');
\path(V2) ++ (200:0.3*\d) coordinate(V2');
\draw (V1)--(V2)--(V3)--(V4)--(V5)--(V1);
\draw (V3)--(V4')--(V5);
\draw (V4)--(V4');
\draw (V3)--(V2')--(V1);
\draw (V2)--(V2');
\node (v1) at (V1) [vertB,label=below:{$v_{1}$}]{};
\node (v2) at (V2) [vertG,label=right:{$v_{2}$}]{};
\node (v3) at (V3) [vertR,label=above:{$v_{3}$}]{};
\node (v4) at (V4) [vertB,label=left:{$v_{4}$}]{};
\node (v5) at (V5) [vertR,label=below
:{$v_{5}$}]{};
\node (v4') at (V4') [vertG,label=right:{$\scriptstyle{v_{4}^{\prime}}$}]{};
\node (v2') at (V2') [vertC,label=left:{$\scriptstyle{v_{2}^{\prime}}$}]{};

\path(Origin) ++ (300:\d) coordinate(G3);
\node (g3) at (G3) []{$G_{3}$};
}

%
%

\newcommand{\Pentagonsquare}
{
\path(3,0) coordinate (Origin);
\path (Origin) ++ (0:0) coordinate(Current);
\foreach \i in{0,1,2,...,4}
{
\path (Current)++(0:0) coordinate(V\i);
\path (Current)++(\i*72: \d) coordinate(Current);
}
\draw (V0)--(V1)--(V2)--(V3)--(V4)--(V0);
\node (v0) at (V0) [vertR]{};
\node (v1) at (V1) [vertB]{};
\node (v2) at (V2) [vertR]{};
\node (v3) at (V3) [vertB]{};
\node (v4) at (V4) [vertG]{};
\path(5,0.3) coordinate (Origin);
\path(Origin) ++(0:0) coordinate(A) ++(0:\d) coordinate(B) ++ (90:\d) coordinate(C) ++ (180:\d) coordinate(D);
\draw (A)--(B)--(C)--(A)--(D)--(B);
\draw(C)--(D);
\draw(V2)--(D);
\node (a) at (A)[vertG,scale=1]{};
\node (b) at (B)[vertR,scale=1]{};
\node (c) at (C)[vertB,scale=1]{};
\node (d) at (D)[vertC,scale=1]{};
}

\maketitle

\begin{abstract}
Interaction between clique number  $\omega(G) $ and chromatic number $\chi(G) $ of a graph is a well studied topic in graph theory. Perfect Graph Theorems are probably the most important results in this direction. Graph $G$ is called \emph{perfect}  if  $\chi(H)=\omega(H)$ for every induced subgraph $H$ of $G$.  The Strong Perfect Graph Theorem (SPGT) states that a graph is perfect if and only if it does not contain an odd hole (or an odd anti-hole) as its induced subgraph. The Weak  Perfect Graph Theorem (WPGT) states that a graph is perfect if and only if its complement is perfect. In this paper, we present a formal framework for verifying these results. We model finite simple graphs in the constructive type theory of Coq Proof Assistant without adding any axiom to it. Finally, we use this framework to present a constructive proof of the Lov\'{a}sz Replication Lemma, which is the central idea in the proof of Weak Perfect Graph Theorem. 
\end{abstract}

\section{Introduction}
\label{sec:intro}
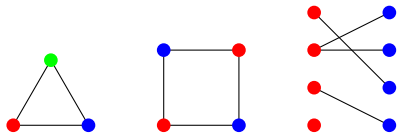
\begin{figure}[H]
\centering
\begin{tikzpicture}
\UTriangle;
\USquare;
\UBipartitegraph;
\end{tikzpicture}
\caption{ \label{fig:F1}Some graphs where  $\chi(G)=\omega(G)$}
\end{figure}

The chromatic number $\chi(G)$ of a graph $G$ is the minimum number of colours needed to colour the vertices so that every two adjacent vertices get distinct colours. Finding out the chromatic number of a graph is  NP-Hard \cite{Garey:1990:CIG:574848}. However, one obvious lower bound is clique number $\omega(G)$, the size of the biggest clique in $G$. Consider the  graphs shown in Figure \ref{fig:F1}.

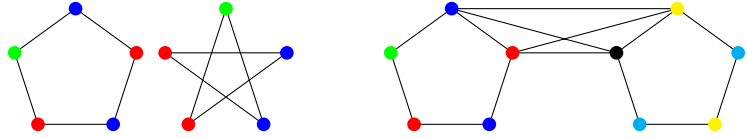
\begin{figure}[H]
\centering
\begin{tikzpicture}
\UPentagon;
\UStar;
\UDoublepentagon;
\end{tikzpicture}
\caption{ \label{fig:F2} Some graphs where, $\chi(G)>\omega(G)$. For $k$ disjoint 5-cycles $\chi(G)= 3k$ and $\omega(G)= 2k$. }
\end{figure}

In each of the above cases \mbox{$\chi(G)=\omega(G)$}, i.e. the number of colours needed  is the minimum we can hope. Can we always hope $\chi(G)=\omega(G)$ for every graph G? The answer is no. Consider the cycle of odd length 5 and its complement shown in Figure \ref{fig:F2}.  In this case one can see that  $\chi(G)=3$ and $\omega(G)=2$ (i.e. $\chi(G)>\omega(G)$).  A cycle of odd length greater than or equal to 5 is called an \emph{odd hole}. Complement of an odd hole is called an \emph{odd anti-hole}. Indeed, the gap between $\chi(G)$ and $\omega(G)$ can be made arbitrarily large. Consider the other graph shown in Figure \ref{fig:F2} which consist of two disjoint 5-cycles with all possible edges between the two cycles. 

This graph is a special case of the general construction where we have $k$ disjoint 5-cycles with all possible edges between any two copies. In this case one can show \cite{ChromaticGap} that $\chi(G)=3k$ but $\omega(G)=2k$. In fact, there is an even stronger result \cite{Mycielski1955}  which constructs triangle-free Micielski graph $M_k$ that satisfies $\chi(M_k)=k$.

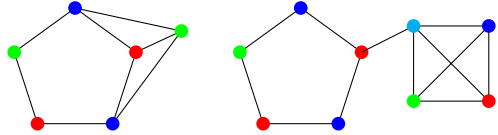
\begin{figure}[H]
\centering
\begin{tikzpicture}
\UExtendedpentagon;
\Pentagonsquare;
\end{tikzpicture}
\caption{ \label{fig:F3} Graphs with $\chi(G)=\omega(G)$,  and  having odd hole as  induced subgraph.}
\end{figure}

In 1961, Claude Berge noticed the presence of odd holes (or odd anti-holes) as induced subgraph in all the graphs presented to him that does not have a nice colouring, i.e. \mbox{$\chi(G)>\omega(G)$}. However,  he also observed some graphs containing odd holes, where $\chi(G)=\omega(G)$.  Consider the graphs shown in Figure \ref{fig:F3}.

A good way to avoid such artificial construction is to make the notion of nice colouring hereditary. We say that a graph $H$ is an induced subgraph of $G$, if $H$ is a subgraph of $G$ and \hbox{$E(H)=\{uv \in E(G) \mid  u,v \in V(H)\}$}. A graph $G$ is then called a \emph{perfect graph} if \mbox{$\chi(H)=\omega(H)$} for all of its induced subgraphs $H$. 

Berge observed that the presence of odd holes (or odd anti-holes) as induced subgraphs is the only possible obstruction for a graph to be perfect. These observations led Berge to the conjecture that a graph is perfect if and only if it does not contain an odd hole (or an odd anti-hole) as its induced subgraph. This was soon known as the Strong Perfect Graph Conjecture (SPGC). Berge thought this conjecture to be a hard goal to prove and  gave a weaker statement referred to as the Weak Perfect Graph Conjecture (WPGC): a graph is perfect if and only if its complement is perfect.  Both  the conjectures are theorems now. In 1972, Lov\'{a}sz proved a result \cite{LOVASZ1972} (known as Lov\'{a}sz Replication Lemma) which finally helped him to prove the WPGC. It took however three more decades to come up with a proof for SPGC. The proof of Strong Perfect Graph Conjecture  was announced  in 2002  by Chudnovsky et al. and finally published \cite{Chudnovsky06thestrong} in  a  178-page paper  in  2006. 

In this paper, we present a formal framework for verifying these results. In Section 2 we provide an overview of the Lov\'{a}sz Replication Lemma which is the key result used in the proof of WPGT.  In Section 3 we present a constructive formalization of finite simple graphs.  In this constructive setting,  we present a formal proof of the Lov\'{a}sz Replication Lemma (Section 4). We summarise the work in Section 5  with an overview of possible future works. The Coq formalization for this paper is available at \cite{list-set}.

\section{Overview of  Lov\'{a}sz Replication Lemma}
\label{sec:LovaszOverview}
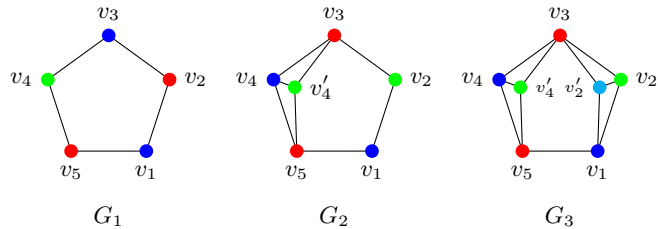
\begin{figure}[H]
\centering
\begin{tikzpicture}
\LPentagon;
\LExtendedpentagon;
\LDextendedpentagon;
\end{tikzpicture}
\caption{ \label{fig:F4} Graph $G_{2}$ is obtained from $G_{1}$ by repeating vertex $v_{4}$ whereas the graph  $G_{3}$ is obtained from $G_{2}$ by repeating vertex $v_{2}$. Note that \mbox{$\chi(G_{2})=\omega(G_{2})=3$} but \mbox{$\chi(G_{3})>\omega(G_{3})$}. }
\end{figure}

Let $G$ be a graph and  $ v \in V(G)$.  We say that $G'$ is obtained from $G$ by repeating vertex $v$ if $G'$ is obtained from $G$ by adding a new vertex $v'$ such that $v'$ is connected to $v$ and to all the neighbours of $v$ in $G$. For example, consider the graphs shown in Figure \ref{fig:F4} obtained by repeating different vertices of a cycle of length 5. 

Note that the graph $G_2$ has a nice coloring (i.e. $\chi(G_{2}) = \omega(G_{2})$), however the graph $G_3$ which is obtained by repeating vertex $v_2$ in  graph $G_2$, does not have a nice coloring (i.e. $\chi(G_{3}) > \omega(G_{3})$).  Thus, property \mbox{$\chi(G)=\omega(G)$} is not preserved by replication. Although, the property $\chi(G)=\omega(G)$ is not preserved,  Lov\'{a}sz in 1971 came up with a surprising result which says that perfectness is preserved by replication. It states that \emph { if $G'$ is obtained from a perfect graph $G$ by replicating a vertex, then $G'$ is perfect}. Note that this result does not apply to any graph shown in Figure \ref{fig:F4}, since none of them is a perfect graph.  All of these graphs  has an induced subgraph (odd hole of length 5) which does not admit a nice coloring. 

\begin{figure}[H]
\centering
\begin{tikzpicture}
\LPolygonABCD;
\LPolygonVaBCD;
\LPolygonVaVbCD;
\LExtendedpolygon;
\end{tikzpicture}
\caption{ \label{fig:F5} The  graphs resulting from repeated replication of vertices $a$, $b$ and $c$  of $G_1$ to form cliques $V_a$ ,$V_b$ and $V_c$ of sizes 2, 3 and 4 respectively. }
\end{figure}
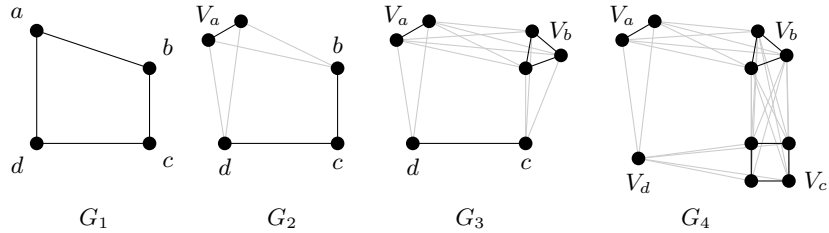

The process of replication can be continued to obtain a graph where  each vertex is replaced with a complete graph of arbitrary size (Figure \ref{fig:F5}).  This gives us a generalised version of the Lov\'{a}sz Replication lemma.

  Let $G$ be a perfect graph and $f:V(G)\rightarrow N$. Let $G'$ be the graph obtained by replacing each vertex $v_{i}$ of the graph $G$ with a complete graph of order $f(v_{i})$. Then $G'$ is a perfect graph.  For example, consider the graphs shown in Figure \ref{fig:F5}.  Vertex $a$ in $G_1$ is replaced by a complete graph $V_a$ of order 2 to obtain $G_2$. Similarly, vertex $b$ in $G_2$ is replaced by a clique $V_b$ of size 3 to obtain $G_3$. Since $G_1$ is perfect, all the other graphs ($G_2$, $G_3$ and $G_4$) obtained by repeated replications are also perfect.

\section{Modeling Finite Simple Graphs}
\label{sec:FiniteGraphs}

There are very few formalization of graphs in Coq. The most extensive among these is due to the formalization of four color theorem \cite{introductionToSsreflect} which considers only planar graphs.  We use a definition for finite graphs which is closest to the one used by Doczkal et. al. \cite{GraphDoc}. However due to reasons explained in this section we represent the vertices of our graphs as sets over \tw{ordType} instead of \tw{eqType}. We define finite simple graphs as a dependent record with six fields. 
\begin{Verbatim} [fontsize=\footnotesize]
Record UG (A:ordType) : Type:= Build_UG {
nodes :> list A;   nodes_IsOrd : IsOrd nodes;    edg: A -> A -> bool;
edg_irefl: irefl edg;   no_edg: edg only_at nodes;    edg_sym: sym edg  }.
Variable G: UG A.       	
\end{Verbatim}
\noindent The last line  in the above code declares a finite graph \tw{G} whose vertices come from an infinite domain \tw{A}. The first field of \tw{G} can be accessed using the term \tw{(nodes G)}. It is a list  that represents the set of vertices of \tw{G}.  The second field of \tw{G} ensures that the  list  of vertices can be considered as a set (details in Sec \ref{subsec:Vertices}). The third field, which is accessed using the term \tw{(edg G)},  is a binary relation  representing the edges of the graph \tw{G}. The terms \tw{(edg\_irefl G)}  and \tw{(edg\_sym G)} are proof terms whose type ensures that the edge relation of \tw{G} is irreflexive and symmetric. These restrictions on edge  makes the graph \tw{G} simple and undirected.  For simplicity in reasoning, the edge relation is considered false everywhere outside the vertices of \tw{G}. This fact  is represented by the proof term \tw{(no\_edg G)}.  

\subsection{Vertices as constructive sets}
\label{subsec:Vertices} 
In our work we only consider finite graphs. Vertices of a finite graph can be represented using a finite set.  The Mathematical Components library \cite{introductionToSsreflect} describes an efficient way of working with  finite sets.  Finite sets are implemented using finite functions (\tw{ffun}) built over a finite type (\tw{finType}). Since all the elements of  a set  now come from a finite domain (i.e. \tw{finType}) almost every property on the set can be represented using computable (boolean) functions. These boolean functions can be used to do case analysis on different  properties of a finite set in a constructive way.

The proof of WPGT involves expansion of graph in which the vertices of the initial graph are replaced with cliques of different sizes. Therefore, in our formalization we can't assume that the vertices of our initial graph are sets over some \tw{finType}.  Instead, we represent the set of vertices of a graph \tw{G} as a list whose  elements  come from an infinite domain (defined as \tw{ordType}).
\subsubsection{Reflection, eqType and ordType }
The \tw{finType} in ssreflect is defined on a base type called \tw{eqType}. The \tw{eqType} is defined as a dependent record which packs together a type  (\tw{T:Type}) and a boolean comparison operator (\tw{eqb: T $\rightarrow$ T $\rightarrow$ bool}) that can be used to check the equality of any two elements of type \tw{T}.   Therefore, it tries to capture the notion of a domain with decidable equality. For example, consider the following canonical instance which  connects natural numbers to the theory of \tw{eqType}.
\begin{Verbatim}[fontsize=\footnotesize]
Canonical nat_eqType: eqType:= 
{|Decidable.E:= nat; Decidable.eqb:= Nat.eqb;  Decidable.eqP:= nat_eqP|}.
\end{Verbatim}
Here, \tw{Nat.eqb} is a boolean function that checks the equality of two natural number and the term  \tw{nat\_eqP} is name of a lemma which ensures that the function \tw{eqb} evaluates  equality in correct way. 
\begin{lemma} 
\tw{ \textcolor{gray}{nat\_eqP} (m n:nat): reflect (m = n)(Nat.eqb m n)}
\end{lemma}
Note, the use of \tw{reflect} predicate to specify a boolean function. It is a common practice in the ssreflect library. Once we connect a proposition \tw{P} with a boolean \tw{B} using the \tw{reflect} predicate we can easily navigate between them. This makes case analysis on \tw{P} possible even though the Excluded Middle principle is not provable for an arbitrary proposition \tw{Q} in the constructive type theory of Coq.  Consider the following lemma  (from  \tw{GenReflect.v}),  which makes case analysis possible on a predicate \tw{P}.
\begin{lemma} 
\tw{ \textcolor{gray}{reflect\_EM} (P: Prop)(b:bool):  reflect P b -> P $\lor$  $\lnot$ P.}
\end{lemma}
To keep the  library constructive we follow this style of proof development.  All the basic predicates on sets are connected with their corresponding boolean functions using  reflection lemmas. For example, consider the predicates mentioned in Table \ref{tab:T1}. 

\begin{table}[h]
\centering
\( \begin{array} {|r|l|l|} \hline
\tw{Propositions (P:Prop)} & \tw{Boolean functions (b:bool) } & \tw{Reflection lemmas} \\ \hline
 \tw{In a l} & \tw{memb a l} & \tw{membP}  \\
 \tw{Equal l s} &  \tw{equal l s} & \tw{equalP} \\
 \tw{$\exists$ x, (In x l $\land$ f x)} & \tw{existsb f l} & \tw{existsbP} \\
 \tw{$\forall$ x, (In x l -> f x)} & \tw{forallb f l} & \tw{forallbP} \\ \hline
\end{array} \)
\caption{ \label{tab:T1} Some decidable predicates on sets from the file  \tw{SetReflect.v}}
\end{table}
These lemmas can be used to do case analysis on any statement about sets containing elements of \tw{eqType}.  However, in this framework we can't be constructive while reasoning about properties of power sets.  Hence, we  base our set theory on \tw{ordType} which is defined as a subtype of \tw{eqType}. 

The \tw{ordType} inherits all the fields of \tw{eqType} and has an extra boolean operator which we call the less than boolean operator (i.e. \tw{ltb}). This new operator represents the notion of ordering among elements of \tw{ordType}. 

\subsubsection{ Sets as Ordered Lists }
Let \tw{T} be an \tw{ordType}. Sets on domain \tw{T} is then defined as a dependent record  with two fields. The first field is a list of elements of type \tw{T} and the second field ensures that the list  is ordered using the \tw{ltb} relation of \tw{T}. 
\begin{Verbatim}[fontsize=\footnotesize]
Record set_on  (T:ordType): Type := { S_of :> list T;  IsOrd_S : IsOrd S_of }. 
\end{Verbatim}

All the basic operations on sets (e.g. union, intersection and set difference) are implemented using functions on ordered lists which outputs an ordered list.  An important consequence of representing sets using ordered list is the following lemma (from \tw{OrdList.v}) which states that the element wise equal sets can be substituted for each other in any context. 
\begin{lemma} 
\tw{ \textcolor{gray}{set\_equal} (A: ordType)(l s: set\_on A): Equal l s ->  l = s. }
\end{lemma}

Another advantage of representing sets using ordered list is that  we can now enumerate all the subsets of a set \tw{S} in a list using the function \tw{pw(S)}.  Moreover, we  have following lemma which states that the list generated by  \tw{pw(S)} is a set. The details of function \tw{pw(S)} can be found in the file \tw{Powerset.v}. 
 \begin{lemma} 
\tw{ \textcolor{gray}{pw\_is\_ord} (S: list A): IsOrd (pw S). }
\end{lemma}

Since all the subsets of \tw{S} are  present in the list \tw{pw(S)} we can  express any predicate on power set using a boolean function on  list. This gives us a constructive framework for reasoning about properties of sets as well as their power sets. For example, consider the following definition of a boolean function \tw{forall\_xyb}  and its corresponding reflection lemma \tw{forall\_xyP}  from  the file \tw{SetReflect.v}.  
\begin{Verbatim}[fontsize=\footnotesize]
  Definition forall_xyb (g:A->A->bool)(l:list A):= 
                 (forallb (fun x=> (forallb (fun y => g x y) l )) l).
  Lemma forall_xyP (g:A->A->bool) (l:list A):
    reflect (forall x y, In x l-> In y l-> g x y)  (forall_xyb g l).               
\end{Verbatim}

\subsection{Decidable Edge relation}
The edges of graph \tw{G} are represented using a decidable binary relation on the vertices of \tw{G}. Hence, one can check the presence of an edge between vertices \tw{u} and \tw{v} by evaluating the expression \tw{(edg G u v)}. The decidability of edge relation is useful for defining many other important properties of graphs as decidable predicates. 
\subsubsection*{Cliques, Stable Set and Graph Coloring}
Consider the following definition of a complete graph \tw{K}  present in the graph \tw{G}. Note that the proposition \tw{Cliq G K} is decidable because it is connected to a term of type bool (i.e. \tw{cliq G K}) by the reflection lemma \tw{cliqP}.  
\begin{Verbatim}[fontsize=\footnotesize]
 Definition cliq(G:UG)(K:list A):=forall_xyb (fun x y=> (x==y) || edg G x y) K.
 Definition Cliq(G:UG)(K:list A):=(forall x y,In x K->In y K-> x=y \/ edg G x y).
 Lemma cliqP (G: UG)(K: list A): reflect (Cliq G K) (cliq G K).                
\end{Verbatim}

In a similar way we also define independence set (or stable set) and graph coloring using decidable predicates. The details can be found in the file \tw{UG.v} and \tw{MoreUG.v}.   Most of these these definitions together with their  reflection lemmas are listed in Table \ref{tab:T2}. 

\begin{table}[h]
\centering
\( \begin{array} {|r|l|l|} \hline
\tw{Propositions (P:Prop)}   &   \tw{Boolean functions (b:bool) }  & \tw{Reflection lemmas} \\ \hline
 \tw{Subgraph G1 G2 }         &   \tw{subgraph G1 G2}                    & \tw{subgraphP}          \\
 \tw{Ind\_Subgraph G1 G2 } &  \tw{ind\_subgraph G1 G2 }            & \tw{ind\_subgraphP}  \\
 \tw{ Stable G I }                    & \tw{stable G I }                                & \tw{stableP}               \\
 \tw{Max\_I\_in G I}                & \tw{max\_I\_in G I}                           & \tw{max\_I\_inP}        \\
 \tw{Cliq G K }                        & \tw{cliq G K}                                    & \tw{cliqP}                   \\
 \tw{Max\_K\_in G K}             & \tw{ max\_K\_in G K}                       & \tw{max\_K\_inP}       \\
 \tw{Coloring\_of G f}           & \tw{coloring\_of G f}                        & \tw{coloring\_ofP}  \\ \hline
\end{array} \)
\caption{\label{tab:T2} Decidable predicates on finite graphs (from \tw{UG.v} and \tw{MoreUG.v}).}
\end{table}
\noindent We call a graph $G$ to be a nice graph if $\chi(G)=\omega(G)$. A graph $G$ is then called a perfect graph if every induced subgraph of it is a nice graph. 
\begin{Verbatim}[fontsize=\footnotesize]
 Definition Nice (G: UG): Prop:= forall n, cliq_num G n -> chrom_num G n.
 Definition Perfect (G: UG): Prop:= forall H, Ind_subgraph H G -> Nice H.
\end{Verbatim}
In this setting we have the following lemma establishing the obvious relationship between $\chi(G)$ and $\omega(G)$. Here the expression \tw{(clrs\_of f G)} represents the set containing all colors used by \tw{f} to color the vertices of \tw{G}.
\begin{lemma} 
\tw{ \textcolor{gray}{more\_clrs\_than\_cliq\_size} (G: UG)(K: list A)(f: A -> nat):
     Cliq\_in G K-> Coloring\_of G f -> |K| <= |clrs\_of f G|. }
\end{lemma}

\begin{lemma} 
\tw{ \textcolor{gray}{more\_clrs\_than\_cliq\_num} (G: UG)(n:nat)(f: A->nat): cliq\_num G n-> Coloring\_of G f -> n <= |clrs\_of f G|.}
\end{lemma}

\subsection{Graph Isomorphism}
It is typically assumed in any proof involving graphs that isomorphic graphs have exactly the same properties. However, in a formal setting we need a proper representation for graph isomorphism to claim the exact behaviour of isomorphic graphs.  
\begin{Verbatim}[fontsize=\footnotesize]
 Definition iso_using (f: A->A)(G G': @UG A) := (forall x, f (f x) = x) /\
   (nodes G') = (img f G)  /\   (forall x y, edg G x y = edg G' (f x) (f y)).               
 Definition iso (G G': @UG A) := exists (f: A->A), iso_using f G G'.
\end{Verbatim}
 Consider the following lemmas which shows the symmetric nature of graph isomorphism.
\begin{lemma} \label{lem: isosym}
\tw{ \textcolor{gray}{iso\_sym} (G G': UG): iso G G' -> iso G' G. }
\end{lemma}

 Note the self invertible nature of \tw{f} which makes it injective on both \tw{G} and \tw{G'}.  The second condition (i.e. \tw{(nodes G') = (img f G)}) expresses the fact that f maps all the vertices of \tw{G} to the vertices of \tw{G'}.  

 \begin{lemma} 
\tw{ \textcolor{gray}{iso\_one\_one} (G G': UG)(f: A-> A): iso\_using f G G'-> one\_one\_on G f. }
\end{lemma}

\begin{lemma} \label{lem: IsoSub}
\tw{ \textcolor{gray}{iso\_subgraphs} (G G' H :UG) (f: A-> A) : iso\_using f G G'-> Ind\_subgraph H G -> ($\exists$ H', Ind\_subgraph H' G' $\land$ iso\_using f H H').}
\end{lemma}    
For the graphs \tw{G} and \tw{G'}   Lemma \ref{lem: IsoSub}  states that every induced subgraph \tw{H} of \tw{G} has an isomorphic counterpart \tw{H'} in \tw{G'}. In a similar way we can prove that the stable sets and cliques in \tw{G} has isomorphic counterparts in  \tw{G'}. For example, consider the following lemmas from \tw{IsoUG.v} summarising these results. 
\begin{lemma} 
\tw{\textcolor{gray}{iso\_cliq} (G G': UG)(f:A-> A)(K:list A):iso\_using f G G'-> Cliq G K -> Cliq G' (img f K).}
\end{lemma}

\begin{lemma} 
\tw{\textcolor{gray}{iso\_stable} (G G': UG)(f: A-> A)(I: list A):iso\_using f G G'-> Stable G I-> Stable G' (img f I).}
\end{lemma}

 \begin{lemma} 
\tw{ \textcolor{gray}{iso\_coloring}(G G':UG)(f:A->A)(C: A->nat):iso\_using f G G' -> Coloring\_of G C -> Coloring\_of G' (fun (x:A) => C (f x)). }
\end{lemma}

 \begin{lemma} 
\tw{ \textcolor{gray}{perfect\_G'} (G G':UG ): iso G G'-> Perfect G -> Perfect G'.}
\end{lemma}

\subsection{Graph constructions}
Adding (or removing) edges in an existing graph to obtain a new graph is a  common procedure in proofs involving graphs. In such circumstances an explicit specification of all the  fields of the new graph becomes a tedious job. 

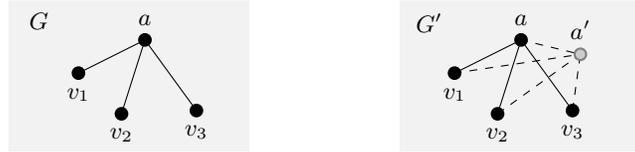
\begin{figure}[H]
\centering
\begin{tikzpicture}[xscale=1, yscale=0.5]
\LGraph;
\end{tikzpicture}
\caption{ \label{fig:F6} Graph $G'$ is obtained from $G$ by repeating vertex $a$ to $a'$. }
\end{figure}

For example, consider the definition of following function \tw{(nw\_edg G a a')}. 
\begin{Verbatim}[fontsize=\footnotesize]
Definition nw_edg(G:UG)(a a':A):= fun(x y:A)=> match (x==a), (y==a') with
                			  | _ , false => (edg G) x y
                			  | true, true => true
                			  |false, true => (edg G) x a
                  		end.
\end{Verbatim}  
The term \tw{(nw\_edg G a a')} can be used to describe the edge relation of graph \tw{G'} shown in Figure \ref{fig:F6}, which is obtained  from \tw{G} by repeating the vertex \tw{a} to \tw{a'}.

This function has a simple definition and hence it is easy to prove various properties about it. For example, we can prove following results establishing connections between the edges  of \tw{G} and \tw{G'}. 
\begin{lemma}
\tw{ \textcolor{gray}{nw\_edg\_xa\_xa'} (G: UG)(a a' x:A): (edg G) x a ->  (nw\_edg G a a') x a'. }
\end{lemma}
\begin{lemma}
\tw{ \textcolor{gray}{nw\_edg\_xy\_xy}  (G: UG)(a a' x y:A)(P': $\lnot$ In a' G): (edg G) x y ->  (nw\_edg G a a') x y}
\end{lemma}
\begin{lemma}
\tw{ \textcolor{gray}{nw\_edg\_xy\_xy4} (G: UG)(a a' x y:A)(P: In a G)(P': $\lnot$In a' G): y $\neq$ a' -> (edg G) x y = (nw\_edg G a a') x y. }
\end{lemma}

 Although, the term \tw{(nw\_edg G a a')} contains  all the essential properties of the construction it doesn't have the irreflexive and symmetric properties necessary for an edge relation. Hence, we can't use this term for edge relation while declaring \tw{G'} as an instance of \tw{UG}. To ensure these properties one can add more branches to the match statement and provide an extra term \tw{P} of type \tw{a  $\neq $ a'} as argument to the function.  However, this will result in a more complex function and proving even the essential properties of the new function becomes hard. 
 
 Instead of writing complex edge relations every time we define functions namely \tw{mk\_irefl}, \tw{mk\_sym} and \tw{E\_res\_to}  which make minimum changes and convert any binary relation on vertices into an edge relation. For example consider the following specification lemmas for the functions \tw{mk\_irefl} and \tw{mk\_sym}. 
 \begin{lemma}
\tw{ \textcolor{gray}{mk\_ireflP} (E: A -> A-> bool): irefl (mk\_irefl E). }
\end{lemma}
 \begin{lemma}
\tw{ \textcolor{gray}{mk\_symP} (E: A-> A-> bool): sym (mk\_sym E). }
\end{lemma}
 \begin{lemma}
\tw{ \textcolor{gray}{irefl\_inv\_for\_mk\_sym} (E: A-> A-> bool): irefl E -> irefl (mk\_sym E). }
\end{lemma}
\begin{lemma}
\tw{ \textcolor{gray}{sym\_inv\_for\_mk\_irefl} (E: A->A-> bool): sym E -> sym (mk\_irefl E).}
\end{lemma}

 Note that these functions do not change the properties ensured by each other. The file \tw{UG.v} contains other invariance results about these functions proving that these functions work well when used together.  For example, consider the following declaration of \tw{G'} as an instance of \tw{UG}. 

\begin{Verbatim}[fontsize=\footnotesize]
 Definition ex_edg(G:UG)(a a':A):=
                             mk_sym(mk_irefl((nw_edg G a a') at_ (add a' G))).
 Variable G: UG.                 
 Definition G':= refine({| nodes:= add a' G; edg:= (ex_edg G a a');
                             |}); unfold ex_edg. all: auto. Defined.
\end{Verbatim} 
 
     Note that the term \tw{(ex\_edg G a a')}  obtained from \tw{(nw\_edg G a a')}  by using these functions have all the properties of an edge relation. Now, we can simply use the tactic \tw{all: auto} to discharge all the proof obligations generated while  declaring  \tw{G'} as an instance of \tw{UG}.  Therefore these functions can significantly ease the construction of new graphs. 
     
     All the important properties of the final edge relation (i.e. \tw{ex\_edg G a a'}) can now be derived from the properties of \tw{nw\_edg G a a'} by using the specification lemmas for the  functions \tw{mk\_irefl}, \tw{mk\_sym} and \tw{E\_res\_to}. For example consider following lemmas (from \tw{Lovasz.v}) which describes the final edge relation (i.e. \tw{ex\_edg}) of graph \tw{G'}.
     \begin{lemma}
     \tw{ \textcolor{gray}{Exy\_E'xy}  (x y:A)(P: In a G)(P': $\lnot$ In a' G): edg G x y -> edg G' x y. }
     \end{lemma}
      \begin{lemma}
     \tw{ \textcolor{gray}{In\_Exy\_eq\_E'xy} (x y:A)(P: In a G)(P': $\lnot$ In a' G): In x G-> In y G-> edg G x y=edg G' x y. }
     \end{lemma}
      \begin{lemma}
     \tw{ \textcolor{gray}{Exy\_eq\_E'xy} (x y:A)(P: In a G)(P': $\lnot$ In a' G): x $\neq$ a'-> y $\neq$ a'->  edg G x y = edg G' x y. }
     \end{lemma}
      \begin{lemma}
     \tw{ \textcolor{gray}{Exa\_eq\_E'xa'} (x:A)(P: In a G)(P': $\lnot$ In a' G): x $\neq$ a-> x $\neq$ a'->  edg G x a = edg G' x a'.}
     \end{lemma}
     \begin{lemma}
     \tw{ \textcolor{gray}{Eay\_eq\_E'a'y} (y:A)(P: In a G)(P': $\lnot$ In a' G): y $\neq$ a -> y $\neq$  a' -> edg G a y = edg G' a' y.}
     \end{lemma}
     
\section{Constructive proof of  Lov\'{a}sz Replication Lemma}
\label{sec:LovaszProof}
Let \tw{G} and \tw{G'} be the graphs discussed in section 3.4, where \tw{G'} is obtained from \tw{G} by repeating the vertex \tw{a} to \tw{a'}. Then we have the following lemma.
\begin{lemma}
\tw{ \textcolor{gray}{ReplicationLemma}   Perfect G -> Perfect G'. }
\end{lemma} 
\textbf{Proof}: We prove this result using induction on the size of graph \tw{G}. 
\begin{itemize}
\item Induction hypothesis(\tw{IH}): \tw{ $\forall$ X, |X|<|G|-> Perfect X -> Perfect X' }
\end{itemize}Let \tw{H'} be an arbitrary induced subgraph of \tw{G'}, then we need to prove that \tw{$\chi$(H') = $\omega$(H')}. We prove this equality in both of the following cases.
\begin{itemize}
\item \textbf{Case-1} (\tw{H' $\neq$ G'}): In this case \tw{H'} is strictly included in \tw{G'}. We further do case analysis on the proposition (\tw{a $\in$ H'}). 
\begin{itemize}
\item \textbf{Case-1A} (\tw{a $\notin$ H'}): In this case if \tw{a' $\notin$ H'} then \tw{H'} is an induced subgraph of \tw{G} and hence \tw{$\chi$(H') = $\omega$(H')}.  Now consider the case when  \tw{a' $\in$ H'}. Let \tw{H} be the induced subgraph of \tw{H'} restricted to the vertex-set \tw{(H'$\setminus$a') $\cup$ \{a\}}. Note that \tw{H'} is isomorphic to \tw{H} and \tw{H} is an induced subgraph of \tw{G}. Hence \tw{H'} is a perfect graph and we have  \tw{$\chi$(H') = $\omega$(H')}. 
\item \textbf{Case-1B} (\tw{a $\in$ H'}): Again in this case if \tw{a' $\notin$ H'} then \tw{H'} is an induced subgraph of \tw{G} and hence \tw{$\chi$(H') = $\omega$(H')}. Now we are in the case where \tw{a $\in$ H'} , \tw{a' $\in$ H'} and \tw{H'} is strictly included in \tw{G'}. Therefore, the set \tw{H'$\setminus$a'} is strictly included in \tw{G}. Let \tw{H} be the induced subgraph of \tw{G} with vertex set \tw{H'$\setminus$a'}. Note that \tw{H'} can be obtained by repeating \tw{a} to \tw{a'} in \tw{H}. But, we know that \tw{ |H| < |G|},  hence \tw{H'} is a perfect graph by induction hypothesis (\tw{IH}) and we have \tw{$\chi$(H') = $\omega$(H')}. 
\end{itemize}
\item \textbf{Case-2} (\tw{H' = G'}): In this case we need to prove \tw{$\chi$(G') = $\omega$(G')}. We further split this case into two sub cases.
\begin{itemize}
\item \textbf{Case-2A}: In this case we assume that there exists a clique \tw{K} of size \tw{$\omega$(G)} such that \tw{a $\in$ K}. Hence \tw{K} gets extended to a clique of size \tw{$\omega$(G)+1} in \tw{G'} and \tw{$\omega$(G')= $\omega$(G)+1}. Now we can assign a new color to the vertex \tw{a'} which is different from all the colors assigned to \tw{G}. Hence \tw{$\chi$(G')= $\chi$(G)+1=$\omega$(G)+1=$\omega$(G')}.
\item \textbf{Case-2B}: In this case we assume that \tw{a} does not belong to any clique \tw{K} of size \tw{$\omega$(G)}. Since \tw{G} is a perfect graph let \tw{f} be a coloring of graph \tw{G} which uses exactly \tw{$\omega$(G)} colors. Let \tw{ G$^*$ = \{ v$\in$ G: f(v) $\neq$ f(a) $\lor$ v=a \}}. For the subgraph \tw{G$^*$} we can then show that \tw{$\omega$(G$^*$) < $\omega$(G)}. Hence there must exist a coloring \tw{f$^*$} which uses \tw{$\omega$(G$^*$)} colors for coloring \tw{G$^*$}. Since \tw{$\omega$(G$^*$) < $\chi$(G)} we can safely assume that \tw{f$^*$}  does not use the color \tw{f (a)} for coloring the vertices of  \tw{G$^*$}. Now consider a coloring \tw{f'} which assigns a vertex \tw{x} color \tw{f$^*$(x)} if \tw{x} belongs to \tw{G$^*$} otherwise \tw{f' (x) = f (a)}. Note that the number of colors used by \tw{f} is at most  \tw{$\omega$(G)}. Hence \tw{$\chi$(G') = $\omega$(G')}. 
\end{itemize}
\end{itemize}
Note that all the cases in the above proof correspond to  predicates on sets and finite graphs. Since we have  decidable representations for all of these predicates, we could do case analysis on them without assuming any axiom. $\square$

\section{ Conclusions and Future Work }
Formal verification of a mathematical theory  can often lead to a deeper understanding of the verified results and hence increases our confidence in the theory. However, the task of formalization soon becomes overwhelming because the length of formal proofs blows up significantly. In such circumstances having a library of facts on commonly occurring mathematical structures can be really helpful.  The main contribution of this paper is a constructive formalisation of finite simple graphs in the Coq proof assistant \cite{StandLib}. This formalization can be used as a framework to verify other important results on finite graphs.  To keep the formalisation constructive we follow a proof style similar to the  small scale reflections technique of the ssreflect. We use small boolean functions in a systematic way to represent various predicates over sets and graphs. These functions together with their specification lemmas can help in avoiding  the use of Excluded-Middle in the proof development. We also describe functions to ease the process of new graph construction. These functions can help in discharging most of the proof obligation generated while creating a new instance of finite graph. Finally, we use this framework to present a fully constructive proof of the Lov\'{a}sz Replication Lemma \cite{LOVASZ1972}, which is the central idea in the proof of Weak Perfect Graph Theorem. One can immediately extend this work by formally verifying Weak Perfect Graph Theorem in the same framework. Another direction of work could be to add in the present framework all the basic classes of graphs and decompositions involved in the proof of Strong Perfect Graph Theorem. This can finally result in a constructive formalisation of strong Perfect Graph Theorem in the Coq proof assistant.

\bibliographystyle{plain} 
\bibliography{wpgt}

\end{document}